\documentclass[manuscript,screen,nonacm]{acmart}
\usepackage[T1]{fontenc}
\usepackage{graphicx}
\usepackage{url}
\usepackage[utf8]{inputenc}
\usepackage{multirow}

\usepackage{xcolor}
\newcommand{\added}[1]{\textcolor{black}{#1}}
\newcommand{\addedtwo}[1]{\textcolor{black}{#1}}

\AtBeginDocument{%
  \providecommand\BibTeX{{%
    \normalfont B\kern-0.5em{\scshape i\kern-0.25em b}\kern-0.8em\TeX}}}

\copyrightyear{2021} 
\acmYear{2021} 
\setcopyright{acmlicensed}\acmConference[MuC '21]{Mensch und Computer 2021}{September 5--8, 2021}{Ingolstadt, Germany}
\acmBooktitle{Mensch und Computer 2021 (MuC '21), September 5--8, 2021, Ingolstadt, Germany}
\acmPrice{15.00}
\acmDOI{10.1145/3473856.3473864}
\acmISBN{978-1-4503-8645-6/21/09}

\begin{document}

\title{Auditing the Biases Enacted by YouTube for Political Topics in Germany}


\author{Hendrik Heuer}
\email{hheuer@uni-bremen.de}
\orcid{0000-0003-1919-9016}
\affiliation{%
 \institution{University of Bremen, Institute for Information Management \& Centre for Media, Communication and Information Research (ZeMKI)}
 \city{Bremen}
 \country{Germany}
 }

\author{Hendrik Hoch}
\email{hhoch@ifib.de}
\affiliation{%
 \institution{Institute for Information Management \& Centre for Media, Communication and Information Research (ZeMKI)}
 \city{Bremen}
 \country{Germany}
 }

\author{Andreas Breiter}
\email{abreiter@uni-bremen.de}
\orcid{0000-0002-0577-8685}
\affiliation{%
 \institution{University of Bremen, Institute for Information Management \& Centre for Media, Communication and Information Research (ZeMKI)}
 \city{Bremen}
 \country{Germany}
 }

\author{Yannis Theocharis}
\email{yannis.theocharis@hfp.tum.de}
\orcid{0000-0001-7209-9669}
\affiliation{%
 \institution{Technical University of Munich, School of Governance/Bavarian School of Public Policy}
 \city{Munich}
 \country{Germany}
 }

\renewcommand{\shortauthors}{Heuer et al.}

\begin{abstract}
With YouTube's growing importance as a news platform, its recommendation system came under increased scrutiny. Recognizing YouTube's recommendation system as a broadcaster of media, we explore the applicability of laws that require broadcasters to give important political, ideological, and social groups adequate opportunity to express themselves in the broadcasted program of the service. We present audits as an important tool to enforce such laws and to ensure that a system operates in the public's interest. To examine whether YouTube is enacting certain biases, we collected video recommendations about political topics by following chains of ten recommendations per video. Our findings suggest that YouTube's recommendation system is enacting important biases. We find that YouTube is recommending increasingly popular but topically unrelated videos. The sadness evoked by the recommended videos decreases while the happiness increases. We discuss the strong popularity bias we identified and analyze the link between the popularity of content and emotions. We also discuss how audits empower researchers and civic hackers to monitor complex machine learning~(ML)-based systems like YouTube's recommendation system.
\end{abstract}



\maketitle

\section{Introduction}

Social networking sites like Facebook and Twitter have become important providers of political news, which increasingly raises questions about their role in aiding citizens to become better informed \cite{VanAelst2017}. Studying the recommendations encountered on platforms like YouTube is timely and urgent \cite{10.1145/3392854,10.1145/3359285,ribeiro2019auditing} because the generation, distribution, and consumption of political information in the new political information environment supported by social media is very different than an environment that is traditionally dominated by the mass media \cite{theocharis2021digitally,HALPERN20131159}. \added{The goal of this investigation is to better understand what recommendations YouTube is making for political topics. Recognizing YouTube as an important provider of news, we discuss ways of making sure that YouTube's recommendations are fair and balanced.} \added{This is important since political information environments have undergone significant changes over the last decade \citep{VanAelst2017}, with the rise of social media as sources for political information leading to new forms of news consumption \citep{Matsa2018}. Social networking sites like Facebook and Twitter have become relevant sources of political news, which increasingly raised questions about their role in aiding citizens to become better-informed \citep{VanAelst2017}. Meanwhile, the political relevance of video-sharing sites like YouTube is still mostly unknown. This creates an important gap in the literature, especially since a politically informed citizenry is one of the cornerstones of a well-functioning democracy \citep{DelliCarpini1997}. It is thus unsurprising that there has been an increasing concern about the role of the ML-based curation system in influencing political opinion and pushing users towards politically extreme content.} 

YouTube \cite{youtube_press_2019} is the second most visited website on the Internet with more than two billion users per month. The popular video-sharing website is also an important news source for a large group of people. 27\% of 80,000+ people from 40 countries stated that they are using YouTube for news \cite{newman_reuters_2020}. According to a YouTube official, 70\% of the videos watched on YouTube are recommended by their machine learning~(ML)-based video recommendation system \cite{solsman_youtubes_2018}. This implies that a large amount of global news consumption is due to YouTube's recommendations. This makes YouTube's recommendation system an important provider of news akin to mass media broadcasting services. Therefore, this paper investigates the biases enacted by YouTube's machine learning (ML)-based recommendation system. YouTube \cite{youtube_creators_home} describes the goal of its recommendation system as matching ``each viewer with the videos they are most likely to watch and enjoy''. Even though YouTube's recommendation system is a media provider that influences 70\% of the videos that 2+ billion users consume, few studies have considered its role in political information production. 

\added{This is surprising because mass media broadcasters} like radio and TV are commonly expected to present news and political content in a fair and balanced manner. In countries like Germany, this is required by laws like the Interstate Broadcasting Agreement. Article 25 of the agreement states that the ``content of private broadcasting must generally indicate a plurality of opinion'' \cite{interstate_broadcasting_agreement}. By law, broadcasters in Germany are expected to give ``important political, ideological and social groups (...) adequate opportunity to express themselves in the full programme services''. The law also mandates that minority views are taken into account. Such rules and regulations do not apply to social media platforms which - despite their importance - remain predominantly an entertainment medium in which political information is a mere by-product.

\added{Considering the large number of YouTube users and the special requirements that German laws have for mass media broadcasters that could be applied to YouTube, it is important to critically examine YouTube's recommendations. In addition to that, YouTube has been accused of spreading fake news \cite{frenkel_facebook_2018,isaac_facebook_2016} and conspiracy theories in general \cite{buzzfeed_las_vegas_2017}. The dangers of filter bubbles and online radicalization have also been mentioned frequently in the context of YouTube \cite{pariser2011filter,Konstan2012}. Our paper is motivated by YouTube's alleged role in the 2018 Chemnitz protests in Germany, where the stabbing of a German citizen by foreigners spawned street demonstrations and rioting. According to the New York Times, a large far-right protest was fueled by YouTube recommendations. Users who wanted to inform themselves about the stabbing were directed towards extremist videos by the ML-based curation system \citep{nytimes_chemnitz_2018}. The Times cites an analysis of Chemnitz-related videos, which suggests that YouTube's recommendations consistently directed users towards ``predominantly conspiracy theorist or far-right'' videos about the incident. The Times quotes the organizer of a local refugee organization from Chemnitz, who observed that: ``When you click on one video, whether you like it or not, another one is proposed that features content from far-right conspiracy theories''. Despite these media reports, there is little systematic research aimed at understanding the recommendations provided by YouTube's ML-based curation system. Theoretical models and past research on media effects during events such as electoral campaigns offer little theoretical or empirical ground for assuming that YouTube could have direct persuasive or mobilizing effects \citep{Kalla2018}. Prior research and newspaper articles have argued that recommendation systems provide increasingly more radical content on a particular topic \cite{nytimes_chemnitz_2018}, thus enacting a filter bubble \cite{pariser2011filter} or acting as an echo chamber \cite{colleoni2014echo,wiki_echo_chamber_2019}. Prior research also highlighted the influence of the emotions evoked by content \cite{doi:10.1080/19312458.2018.1479843,stieglitz2013emotions,ferrara2015quantifying}.}

\addedtwo{Motivated by the scenario the New York Times described for the incidents in Chemnitz, we developed a bot to simulate users who want to inform themselves about a new political topic. Using the bot, we collected YouTube recommendations for political topics in Germany. Informed by previous research \cite{stieglitz2013emotions,ferrara2015quantifying}, we expected recommendations for political topics to zoom in on obscure or fringe topics. We also expected recommendations to become less popular and more niche over time. This could be indicative of a filter bubble. In addition to that, reports \cite{doi:10.1080/19312458.2018.1479843,roose_youtubes_2019,silverman_viral_2016} indicates that negative emotions play a central role in online radicalization, filter bubbles, and algorithmic bias. This raises the question of whether recommendations evoking strong negative emotions are recommended by YouTube. For these reasons, we aimed to explore the following research questions:}

\begin{itemize}
  \item \added{\textbf{RQ1}: How does the popularity of recommended videos as measured by views and likes change between recommendations?}
  \item \added{\textbf{RQ2}: Do the recommendations stay on topic or can a topic drift be observed?}
  \item \added{\textbf{RQ3}: How does the emotional content of the videos change between recommendations?}
\end{itemize}

\addedtwo{Our findings indicate that YouTube’s recommendations enact a strong popularity bias (RQ1) and a noticeable emotionality bias (RQ3), without zooming in on a particular topic (RQ2). This means that YouTube has a tendency to recommend content that aligns with the interests of the majority, thus making specific content popular. We find that YouTube's recommendations tend to favor the popular, not the extreme.} Considering reports on how extremist groups use video-posting and keywords \cite{marwick2017media,lewis2018alternative}, we discuss the implications of the popularity bias. In addition to that, the paper makes an important methodological contribution by showing how audits enable such investigations. The paper analyzes the potential biases enacted by YouTube's recommendation system using a sock puppet audit based on random walks \cite{sandvig2014auditing}. Each random walk consists of an initial video and a chain of ten recommendations. For each video in the chain, a Firefox-based bot randomly selected one of the top ten video recommendations displayed in the right sidebar next to the video. To understand how popularity affects the recommendations, we analyzed how metrics like the number of views and likes changed between the initial videos and the recommendations. We also performed an in-depth qualitative analysis of a subset of videos to examine how closely related the follow-up recommendations are to the topics initially entered into the search bar. Motivated by existing research on the emotional impact of the ways in which news is framed, as well as the well-studied and important consequences of how it impacts opinion formation \cite{Kuhne2015}, we also examined whether the videos consistently evoke certain emotions. 

\added{In contrast to our assumptions, our investigation discovered an important topic drift (RQ2). Even though we used political topics as a starting point, the recommendations presented to users are not related to political topics. We found that YouTube's recommendation system was pushing increasingly more popular content as measured by the number of views and likes (RQ1). We also found that the sadness evoked by the videos decreased significantly, while the happiness increased (RQ3).} The paper provides explanations for these findings and discusses their implications for the process of political opinion formation. The methodological contribution of this paper is a description of how audits can be used to monitor complex ML-based systems and to enforce laws like the German Interstate Broadcasting Agreement.

\section{Background}

Recommendation systems, which were initially used to find similar content like movies and songs, are progressively applied to select and rank news based on some criteria of relevance and in regards to limitations of time and space \cite{eslami_i_2015}. \added{Thus, recommendation algorithms are increasingly acting as public relevance algorithms \cite{doi:10.1177/1461444809342738}}. They select and exclude information and define what is considered legitimate or relevant knowledge \cite{doi:10.1177/1461444809342738}. This production of calculated publics potentially can provoke changes in users' behavior and their practices. 

\subsection{Algorithmic Experience}

Machine Learning is increasingly applied in socio-technical systems like YouTube. Even though these systems organize, select and present information, the understanding of algorithmic curation is limited \cite{rader_understanding_2015,eslami_i_2015}. The prevalence of such opaque and ML-based systems has led to a variety of investigations of the user awareness of algorithmic curation, as work by Eslami et al. \cite{eslami_i_2015}, Rader and Gray \cite{rader_understanding_2015}, and Alvarado et al. \cite{alvarado2020youtube} proves. While YouTube's published research describes the general idea of their recommendation system in a paper by Covington, Adams \& Sargin \cite{covington2016deep}, it remains unclear how the system works and what factors it takes into account. Despite the growing importance of recommendation systems that are based on machine learning~(ML), designers and developers' understanding of ML and its applications is only emerging \cite{Dove:2017:UDI:3025453.3025739,doi:10.1177/20539517211017593}. 

Algorithmic transparency is an important and timely concern. Eslami et al. \cite{eslami_i_2015} showed that a majority~(62.5\%) of users is not aware of the existence of algorithms like Facebook's News Feed. Eslami et al. also showed that users are upset when posts by close friends and family members are not shown in their feeds. However, users mistakenly believe that their friends intentionally chose not to show them these posts. This suggests that a lack of awareness can have negative effects on the lives and relationships of users. Rader \& Gray \cite{rader_understanding_2015} investigated how well users understand Facebook's News Feed. User beliefs about Facebook's News Feed ranged from privacy concerns over consumer preferences to speculations about an algorithm that prioritizes posts. Alvarado et al. \cite{alvarado2020youtube}  presented a similar investigation about YouTube's ML-based recommendation system. They found that even users without a background in technology have an intuitive grasp of the socio-technical system around ML-based curation systems.

Alvarado \& Waern \cite{alvarado_towards_2018} argue that the interaction with and experience of algorithms should be made explicit in social media contexts. To enable this, they distinguish five categories of AX: 1. algorithmic profiling transparency, 2. algorithmic profiling management, 3. algorithmic user-control, 4. selective algorithmic memory, and 5. algorithmic awareness. This paper contributes to algorithmic awareness by helping researchers and users understand YouTube's recommendations. This knowledge about what the system does can be used to inform algorithmic user control.

Schou \& Farkas \cite{schou2016algorithms} associate recommendation systems on social media platforms like Facebook with important epistemological challenges. They pose the question of how the potential hidden agendas of social network sites and their role in pre-selecting what appears as representable information can be analyzed. They argue that a website like Facebook is a complex socio-technical network with human and non-human actors, who all influence how information is accessed and understood. \added{This paper extends on this prior work by showing novel methodological approaches for studying algorithmic systems in detail and without privileged access.}

\added{The work presented in this paper also extends on prior work that explores the biases enacted by ML-based systems}. Prior work documented how such systems discriminate against people based on gender, ethnicity, marital, or health status \cite{10.1145/2939672.2945386}. \added{Using a Firefox-based bot that we developed, we show what algorithmic biases YouTube's recommendation systems enact}.

\subsection{Online Disinformation}

\added{This paper investigates the experiences of mass media users in relation to political issues. We extend on prior work focused on Twitter, which explored political issues like political orientation \cite{colleoni2014echo} or political alignment \cite{conover2011predicting}}. In this context, we focus on YouTube and online disinformation, which is an important issue that has recently become the focus of scientific interest. Among others, Marwick \& Lewis \cite{marwick2017media} investigate media manipulation and disinformation on platforms like YouTube. They describe how media manipulation could contribute to a decreased trust in mainstream media, increased misinformation, and the radicalization of users. Their report describes how far-right groups use social media, memes, and bots to increase the visibility of their ideas. Lewis \cite{lewis2018alternative} argues that ``alternative'' influencers adopt techniques of brand influences like relatability, authenticity, and accountability. She discusses how such influencers leverage certain affordances of YouTube as a platform to ``sell'' far-right ideology. She highlights the dangers that this poses for vulnerable and underrepresented populations like the LGBTQ community, women, immigrants, and people of color. This motivated our investigation of what YouTube is recommending to users and whether certain biases are enacted.

Such bias has been conceptualized as an ``imbalance or inequality of coverage rather than as a departure from truth'' \cite[p.~213]{stevenson1973untwisting}, ``any systematic slant favoring one candidate or ideology over another'' \cite[p.~302]{waldman1998newspaper}, or ``the consistent patterns in the framing of mediated communication that promote the influence of one side in conflicts over the use of government power.'' \cite[p.~166]{entman2007framing}. 

Considering these definitions and the diversity of content circulating on YouTube, we operationalize bias in the context of YouTube's recommendations as: 

\begin{quote}
An inclination, prejudice, or overrepresentation for or against one person, group, topic, idea, or content, especially in a way considered to be unfair.
\end{quote}

The scientific study of the selective exposure to information goes back at least to the 1960s. A review of the term selective exposure by Sears \& Freedmann \cite{sears1967selective} found that it is used to refer to ``any systematic bias in audience composition'' and ``unusual agreement about a matter of opinion''. The review suggests that people are disproportionately exposed to communications that support their opinions. Overall, they find that mass media exposes people to views that they are already sympathetic to. 

\subsection{YouTube Recommendation \& Topic Drift}

\added{Due to their impact on video consumption, YouTube recommendations are increasingly studied by scholars \cite{10.1145/3392854,10.1145/3359285,ribeiro2019auditing}, especially considering challenges like disinformation and fake news \cite{lazer_science_2018,10.1145/3240167.3240172}}. The idea that social media exposes people to views that they are already sympathetic to relates to concepts such as echo chambers and filter bubbles. Prior research suggests that YouTube's recommendations can enact an ``ideological bubble'' where users who accessed videos about the extreme right receive recommendations of more extreme right content \cite{doi:10.1177/0894439314555329}. 

However, recent research suggests that this tendency to promote increasingly extreme content may be better controlled now. An analysis by Ribeiro, Ottoni, West, Almeida \& Meira \cite{ribeiro2019auditing} suggests that YouTube is not recommending extreme right videos to politically right-leaning viewers. Meanwhile, their analysis of user comments suggests that users do migrate from milder to more extreme content over time. This is supported by Ledwich \& Zaitsev \cite{ledwich2019algorithmic}, who find that YouTube's recommendation system actively discourages users from visiting extremist content. Their analysis suggests that YouTube directs traffic towards the two largest mainstream groups – the Partisan Right and the Partisan Left – away from more niche content they labeled Conspiracy, White Identitarian, and Anti-Social Justice Warrior. This raises the question whether YouTube's recommendations remain related to a particular topic or whether a topic drift can be observed and whether YouTube's recommendation system is enacting certain biases. \added{Our work extends on this prior work by also investigating the role of emotions like happiness and sadness and by discussing the implications of algorithmic bias in relation to German laws like the Interstate Broadcasting Agreement.}

\subsection{Bias \& Information Systems}

In light of our interest in imbalances in the presentation of certain content on YouTube, the following section provides a cursory overview of work related to biases in information systems. Sweeney \cite{10.1145/2447976.2447990}, for instance, exposed discrimination in online ad delivery by Google. She found that names frequently given to African-American babies led to ads suggestive of an arrest record. An example of discrimination based on gender is provided by Datta, Tschantz \& Datta \cite{AutomatedExperimentsonAdPrivacySettings}, who found that those whose gender was set to female are receiving fewer instances of ads related to high paying jobs. Jannach, Lerche, Kamehkhosh \& Jugovac \cite{jannach2015recommenders} show that recommendation systems regularly favor already popular items. They showed that such popularity biases exist for content like movies, books, hotels, as well as mobile games. According to their analysis, the popularity bias is due to a strong reliance on accuracy metrics like mean absolute error (MAE), root-mean-square error (RMSE), precision (i.e. the positive predictive value), and recall (i.e. sensitivity), which lead systems to focus on a tiny fraction of the item spectrum. They discuss possible strategies to deal with such biases. These strategies include exploring algorithmic alternatives, relying on multiple metrics, and balancing existing trade-offs between accuracy and catalog coverage. Such popularity biases can also be observed on other platforms. Boratto, Fenu \& Marras \cite{boratto2019effect}, for instance, discovered a popularity bias in the context of massive open online courses. They find that a popularity bias in education can lead to a market of courses that is dominated by a few teachers. 

\subsection{Communication and Emotional Contagion}

\added{Motivated by reports that negative emotions play a central role in algorithmic bias and filter bubbles \cite{doi:10.1080/19312458.2018.1479843,roose_youtubes_2019,silverman_viral_2016}, we also examine the connection between recommendation systems and emotional contagion.} According to Cosley, Lam, Albert, Konstan \& Riedl \cite{cosley_is_2003}, the psychological literature on conformity suggests that a system that helps people make choices affects people's opinions. This is especially problematic in the context of political topics. Epstein \& Robertson \cite{EpsteinE4512} showed that biased search engine results can shift the voting preferences of undecided voters by 20\% or more. Even worse, the shift can be much higher in some demographic groups and the search ranking bias can be masked so that people are not aware of the manipulation.

A long tradition in political psychology has established a strong link between emotion and cognition, showing that different emotions have different attitudinal and behavioral consequences -- many of which are democratically useful \cite{roseman1991}. Emotions like enthusiasm and fear, for example, have been found to encourage public attention to politics and motivate involvement in political affairs and election campaigns \cite{marcusetal2017}. When it comes to digital media, prior investigations were based on the emotions felt by the participants in response to multiple videos. Prior work on the communication of emotions by Derks, Fischer \& Bos \cite{DERKS2008766} suggests that computer-mediated communication is as emotional and as personal as face-to-face communication. While online and offline communication are similar, emotion communication is more frequent and more explicit when mediated through computers. 

Motivated by this prior research, this paper investigates the role emotions play in video recommendations. Kramer, Guillory \& Hancock \cite{Kramer8788} investigated emotional contagion on Facebook. Their results imply a large potential for strong effects by online social networks. Kramer et al. show that reducing the amount of content with positive emotions leads people to produce less positive content. Increasing the amount of positive content makes people produce more positive content. The results indicate that emotions shared on Facebook are influenced by the emotions encountered on Facebook. Lee \cite{doi:10.1177/1461444811419829} studied emotional expressions on YouTube in the context of the death of Michael Jackson. She uncovered emotions like sadness, grief, anger, and frustration in user comments. Lee analyses the important role that YouTube plays in facilitating emotional expressions and shows how users can depend on content provided by YouTube to meet their emotional needs, extending on the Media System Dependency theory by Ball-Rokeach \& DeFleur \cite{doi:10.1177/009365027600300101}, which predicts that the emotional response of individuals on their environment is changed by media consumption. Ball-Rokeach and DeFleur further argue that media information resources are a key condition for the alteration of audience beliefs and behavior.

\subsection{Auditing Algorithms}

The audits explored in this paper are motivated by Sandvig, Hamilton, Karahalios \& Langbort \cite{sandvig2014auditing}, who distinguish between five different kinds of algorithm audits:

\begin{enumerate}
\item Code audits entail obtaining a copy of an algorithm and studying the instructions in detail. Unfortunately, code audits are not suitable in the context of YouTube because ML-based curation systems rely heavily on data.
\item Noninvasive user audits examine interactions with a platform using a survey format, where users are asked about a certain platform. Such audits are not well-suited due to the potential biases in the self-reported data.
\item Scraping audits query a particular URL and obtain a large number of data points. However, they are not well-equipped for interactive platforms.
\item Sock puppet audits are based on a computer program that impersonates a user, simulating real usage of a platform while interacting with a system. 
\item Crowdsourced audits / collaborative audits require a large number of people that use a particular platform to gather data, e.g. through a platform like Amazon Mechanical Turk. 
\end{enumerate}

Considering the complexity of recruiting and coordinating a crowdsourced audit, we relied on a sock puppet audit to systematically gather a large number of videos based on a representative interaction with an ML-based curation system.

This paper is focused on political topics because citizens' political information consumption has important consequences for opinion formation and, ultimately, for democratic health \cite{DelliCarpini1997}. Since the political relevance of video-sharing sites like YouTube is still largely unknown, an important gap exists in the literature. The gap is especially important considering that a politically informed citizenry is one of the cornerstones of a well-functioning democracy \cite{DelliCarpini1997}.

This focus is informed by Crawford \cite{crawford_can_2015}, who examined the kind of politics that algorithms instantiate. She characterizes platforms like YouTube as ``highly contested online spaces of public discourse'' and problematizes the role of algorithms in producing clear winners. 

Harcup \& O'Neill \cite{doi:10.1080/1461670X.2016.1150193} examined the values within mainstream journalism. They identified criteria such as recency, conflict, unexpectedness, relevance, proximity, and social impact as values and highlights the potential problem of algorithms recentering public engagement around the complementary interests of the broad majority and profitability. According to Harcup and O'Neill, algorithmic systems can lead to a populist ``profitable and normal'' media experience. This motivated us to systematically investigate the recommendations for relevant search queries on political topics. This potential problem directly relates to the popularity bias that this paper identifies for YouTube's recommendations for political topics in Germany.

This paper discusses audits of the machine learning-based video curation system employed by YouTube. Until recently, the recommendation system on YouTube has received comparatively little attention \cite{alvarado2020youtube}. Our work extends on an investigation by Hussein, Juneja \& Mitra \cite{10.1145/3392854}, who performed audits to understand whether personalization (based on age, gender, geolocation, or watch history) contributes to amplifying misinformation on YouTube. The large-scale, quantitative investigation of YouTube search results, Up-Next recommendations, and Top 5 recommendations by Hussein et al. \cite{10.1145/3392854} finds that demographics, such as gender, age, and geolocation do not have a significant effect on amplifying misinformation in returned search results for users with brand new accounts. Unlike Robertson, Jiang, Joseph, Friedland, Lazer \& Wilson \cite{10.1145/3274417}, their results indicate the existence of a ``filter bubble'' effect. Watching videos that promote misinformation leads to more misinformative video recommendations. This paper extends on their findings and shows that even for the very specific case of German political topics, YouTube recommendations become significantly more popular measured by views and likes and significantly less related to political topics.

Until now, audits have not been used to systematically study political topics on YouTube, especially not in Germany and especially not regarding important political events. Previous research by Wu, Rizoiu \& Xie \cite{10.1145/3359285} employed audits to study YouTube's recommendations for music videos. Their large-scale investigation of a YouTube video network with 60,740 music videos revealed that recommendations focus on a small number of popular videos on YouTube. 82.6\% of the views of music videos on YouTube were based on videos that were recommended next to the most popular videos. Wu et al. \cite{10.1145/3359285} propose a model that allows predicting video popularity, which consistently outperforms baselines. Our paper extends on their findings by focusing on political topics from Germany for a real-world example based on keywords derived from a representative poll in Germany.

\section{Methods}

In this paper, we use audits to study YouTube's recommendations. We present a quantitative investigation that examines YouTube's recommendations in regards to their popularity, the topics they cover, and their emotional content. Due to their societal importance, we focused on political topics. We selected nine political topics from a representative telephone poll conducted on behalf of a German public broadcasting service \cite{wdr_ard-deutschlandtrend_2018}. The topics included the most pressing issues for German citizens at the time. The topics are 1. asylum and refugees, 2. the trade conflict with the USA, 3. the impact of digitalization, 4. protection against crime, 5. climate change and the energy transformation, 6. social policy (e.g. the development of pensions), 7. the creation of affordable housing, 8. school and education policy, and 9. the situation in elderly care. 

We perform a sock puppet audit using the Random Walk method, which has been previously applied to study YouTube \cite{pewresearch_youtube_2018}. We performed 150 random walks that always followed the same procedure. The Firefox-based bot: 

\begin{enumerate}
\item randomly picked one of the nine political topics from Germany, 
\item entered the topic into the YouTube search bar, 
\item randomly picked one of the top ten search results, 
\item saved the video page and watched it for a random number of seconds, 
\item randomly selected one of the top ten video recommendations displayed in the right sidebar next to the video, and 
\item repeated this procedure ten times. 
\end{enumerate}

This allowed the bot to obtain a large number of video recommendations. Overall, we collected 1,650 videos in 150 random walks, including 150 initial videos based on the search results and 150 videos at each step in the chain of recommendations. Each random walk is a representative simulation of a user session that would have lasted several hours and in which a user searches for a particular political topic for the first time. We collected between 12 and 25 random walks per topic. For each random walk and each topic, we started a new browser instance and cleared all cookies. All random walks were collected in May 2019 with the same laptop, on the same network, and with the same IP address. The investigation was conducted from the network of a large campus university in a large city in Germany.

We used all 1,650 videos to answer RQ1 regarding the popularity of the content. To investigate whether the content of the videos changed, three independent raters~(one male, two female) rated a subset of the videos. \added{Since the rating of emotions in videos is a challenging task, we merely used the ratings for a relative comparison of different videos. The raters were recruited from a pool of research assistants at a large campus university in Germany. The raters had a scientific background in media and communication studies (two) or social science (one). All raters were in their early to late twenties. The rating task had no time limit. Raters were paid by the hour. They were trained individually for the task of rating the emotions in the videos. Raters were not aware of the research questions of the investigation. In the training, we discussed and explained the rating criteria and the 11-point Likert scale in person and answered any questions.}

The content analysis was required to answer RQ2 and RQ3. For the content analysis, we randomly selected three random walks for each of the nine topics and coded three videos per random walk: the initial video, the 5th recommendation, and the 10th recommendation. The decision to select the 5th and the 10th recommendation for the in-depth analysis was made at the beginning of the study, i.e. before we reviewed any of the material and before we performed any kind of analysis. The raters reviewed all videos in the same randomized order. They were not aware of the research questions and did not know about the goals of the investigation. Each rater was required to watch at least five minutes of each video before making his or her decision to get a good idea about a video's content and valence. The raters also assessed how closely related the videos are to the political topics. For each video, they had to review the list of topics, find the topic that the video was most related to and then rate how related they thought the video was to that topic. The scale ranged from \textit{``not related at all''}~(0) to \textit{``very related''}~(10). For each video, the raters were also asked to ``Please tell us how much you feel each of the following emotions while watching the video''. They rated whether the videos evoked sadness or happiness on an 11-point Likert scale from \textit{``least''}~(0) to \textit{``most''}~(10). For the comparison, we relied on the mean ratings of all three raters as a measure of central tendency. To investigate whether the number of views and likes changed between the recommendations, we performed nonparametric, two-tailed Mann-Whitney U tests to compare the independent samples. We used Mann-Whitney since our data does not follow a normal distribution and since we did not want to make assumptions about how to interpret the differences in ranks. We checked the inter-rater agreement for the in-depth analysis by computing Krippendorff's alpha for our ordinal, not normally distributed data.

\section{Results}

In the following, we will report the results of our exemplary audit of YouTube recommendations for political topics. First, we will consider the popularity of the recommended videos. After that, we will describe how we analyzed the content in the videos. Based on this analysis, we will investigate the topic drift of the recommendations. Finally, we will explore the emotional content of the videos.

\subsection{Popularity of Recommended Videos}

\begin{figure}
\centering
 \includegraphics[width=.6\columnwidth]{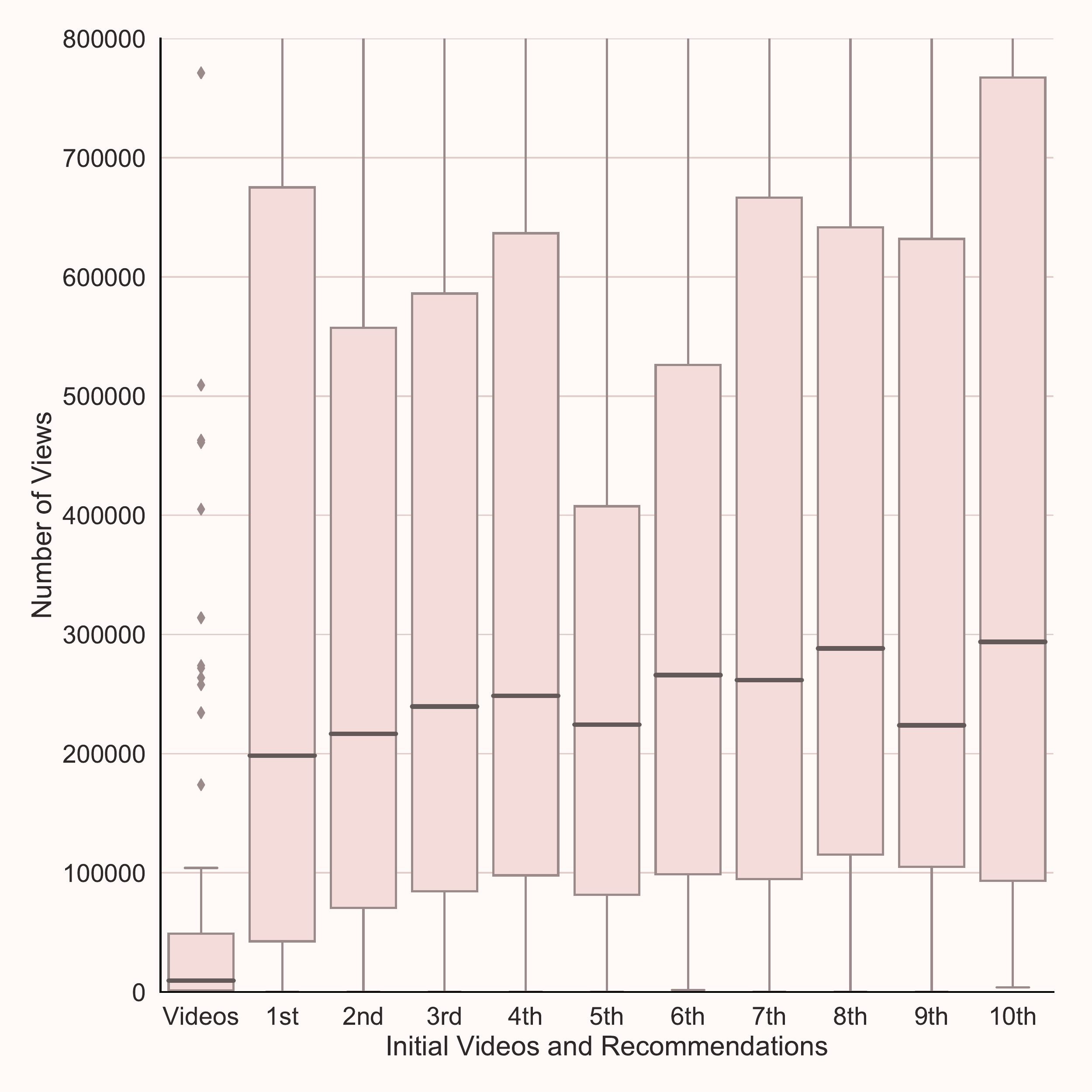}
 \caption{The boxplots show the number of views of the initial videos and the 1st to 10th recommendations}\label{fig:popularity_views}
\end{figure}

\begin{figure}
\centering
 \includegraphics[width=.6\columnwidth]{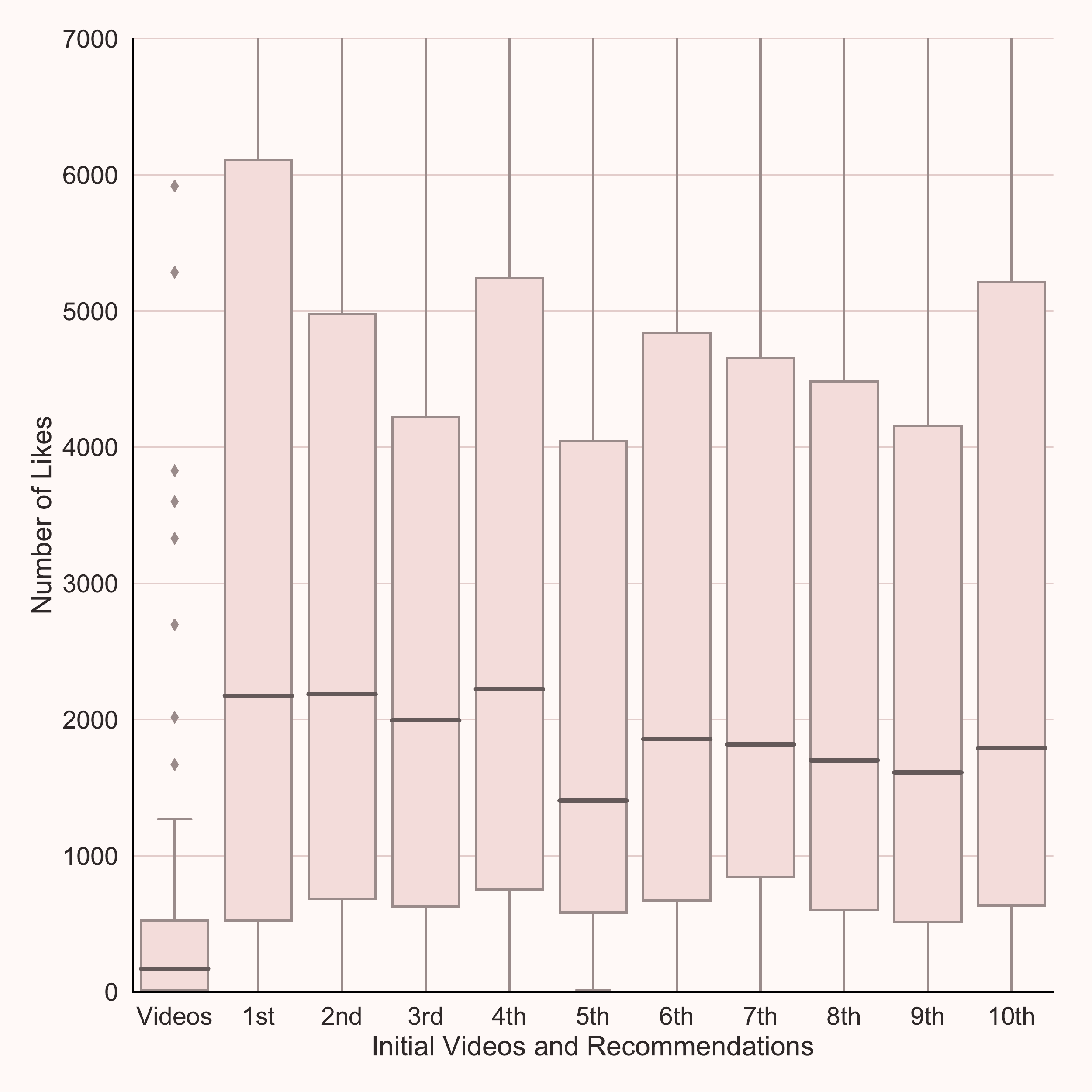}
 \caption{The boxplots show the number of likes of the initial videos and the 1st to 10th recommendations}\label{fig:popularity_likes}
\end{figure}

For RQ1, we investigate whether the popularity of recommended videos changes between recommendations, i.e. whether the 5th recommendations are more popular than the initial videos and whether the 10th recommendations are more popular than the 5th recommendations. For this, we operationalized popularity as the number of views and likes. We included both because views are an implicit measure of popularity while likes are an explicit measure of popularity. Regarding views, it also remains unclear how many seconds a video must be watched before it counts as a view.

We found that the recommendations become significantly more popular. Figure~\ref{fig:popularity_views} shows boxplots of the views of the initial videos and the n-th recommendations, Figure~\ref{fig:popularity_likes} shows boxplots of the likes. For both, a steep increase from the initial videos to the recommendations can be observed. The boxplots show that the median number of views is increasing the longer recommendations are followed. Table~\ref{tab:popularity_statistics} provides the median and mean number of views and likes. Comparing the initial videos and the 5th recommendations, a strong increase in views and likes can be observed, especially between the initial videos and the 5th recommendations. While the initial videos have a median of 9,500 views, the 1st recommendations have a median of around 200,000 views~(Figure~\ref{fig:popularity_views}). After following a chain of ten recommendations, the videos have a median of almost 300,000 views. The number of likes also increases significantly. The initial videos have a median of 170 likes, while the 5th recommendations have a median of 1,404 likes. This further increases to over 1,700 for the 10th recommendations. The finding that the number of views and likes changes between the initial videos and the recommendations is supported by two-tailed Mann–Whitney U tests. The results in Table~\ref{tab:popularity_mannwhitneyu} show statistically significant differences between the initial videos and the 5th recommendations as well as the initial videos and the 10th recommendations.

\begin{table}
\centering
\caption{The median~(Mdn.), mean~($\overline{X}$), and standard deviation~($\sigma$) of views, likes, and channel subscribers for the initial videos~(N=150), as well as the 5th~(N=150), 10th~(N=150), and all recommendations~(N=1,650).}~\label{tab:popularity_statistics}
\begin{tabular}{c c r r r r}
& & & \multicolumn{3}{c}{\textbf{Recommendations}} \\\cmidrule(r){4-6}
\textit{Metric} & \textit{} & \textit{Video} & \textit{5th} & \textit{10th} & \textit{Overall} \\
\midrule
\multirow{3}{*}{Views}  & Mdn.  & 9,590 & 224,353 & 293,789 & 249,754 \\
  & $\overline{X}$ & 93,879  & 467,457 & 838,232 & 602,292 \\
  & $\sigma$ & 212,762 & 820,982 & 2,005,223 & 1,681,245 \\[.25\normalbaselineskip]
\multirow{3}{*}{Likes}  & Mdn.  & 170 & 1,404 & 1,788 & 1,852 \\
  & $\overline{X}$ & 2,213 & 4,143 & 7,998 & 5,328 \\
  & $\sigma$ & 6,823 & 10,664  & 31,183  & 16,198  \\[.25\normalbaselineskip]
\end{tabular}
\end{table}

\begin{table}
\centering
\caption{Two-tailed Mann–Whitney U tests confirm significant differences between the views and likes of the initial videos and the recommendations~($p<.0001$).}~\label{tab:popularity_mannwhitneyu}
\begin{tabular}{r c c r l}
& & & \multicolumn{2}{c}{\textbf{Mann-Whitney U tests}} \\\cmidrule(r){4-5}
\textit{Metric} & \multicolumn{2}{c}{\textit{Comparison between}} & \textit{U} & \textit{p} \\
\midrule 
\multirow{3}{*}{Views} & Video & 5th Rec. & 3728.5 & 0.0000 *** \\
& Video & 10th Rec. & 3167.0 & 0.0000 *** \\
& 5th Rec. & 10th Rec. & 9819.5 & 0.0570   \\[.25\normalbaselineskip]
\multirow{3}{*}{Likes} & Video & 5th Rec. & 4796.5 & 0.0000 *** \\
& Video & 10th Rec. & 4874.0 & 0.0000 *** \\
& 5th Rec. & 10th Rec. & 10471.0 & 0.3001 \\[.25\normalbaselineskip]
\end{tabular}
\end{table}

\subsection{Content Analysis of Videos}

For this study, we also investigated the content of the videos. The goal of our investigation was to compare the relative change in topics, and the emotions evoked by the videos. To investigate this, we relied on ratings provided by independent raters. The three raters reviewed 76 videos, not 81 videos because five videos were deleted between the time we performed the random walks and the time the raters reviewed the videos. 

We computed Krippendorff's alpha coefficient~($\alpha$) for ordinal data to get some indication of how much the raters agreed. Krippendorff's alpha is a generalization of several inter-rater agreement statistics. When discussing the inter-rater agreement, we refer to the terminology by Landis \& Koch \cite{10.2307/2529310}. Based on Krippendorff's alpha, we found substantial agreement regarding how similar the videos were to the topics in our investigation~(.765) and for the sadness evoked by the videos~(.613). We found a moderate agreement for the happiness~(.441) of the videos. 

To put the inter-rater agreement into perspective, we compare our results to prior work. Emotion recognition is a challenging problem for humans and machine learning systems alike, especially in videos \cite{Dhall:2015:VIB:2818346.2829994}. Inter-rater agreement for affective content analysis in videos is expected to be considerably lower than the agreement expected for other coding tasks in the social sciences. For instance, Abrilian et al. report an inter-rater agreement as measured by Cronbach's alpha of 0.254 for intensity and 0.574 for valence of French video clips~\cite{abrilian2005emotv1}. Baveye, Dellandréa, Chamaret \& Chen \cite{7024148} report a Krippendorff's alpha of 0.191 for arousal and 0.180 for valence regarding their video database LIRIS-ACCEDE. We, therefore, conclude that the inter-rater agreement is sufficient for a relative comparison.

\subsection{Topic Drift of Recommendations}

\begin{figure}
\centering
 \includegraphics[width=.35\columnwidth]{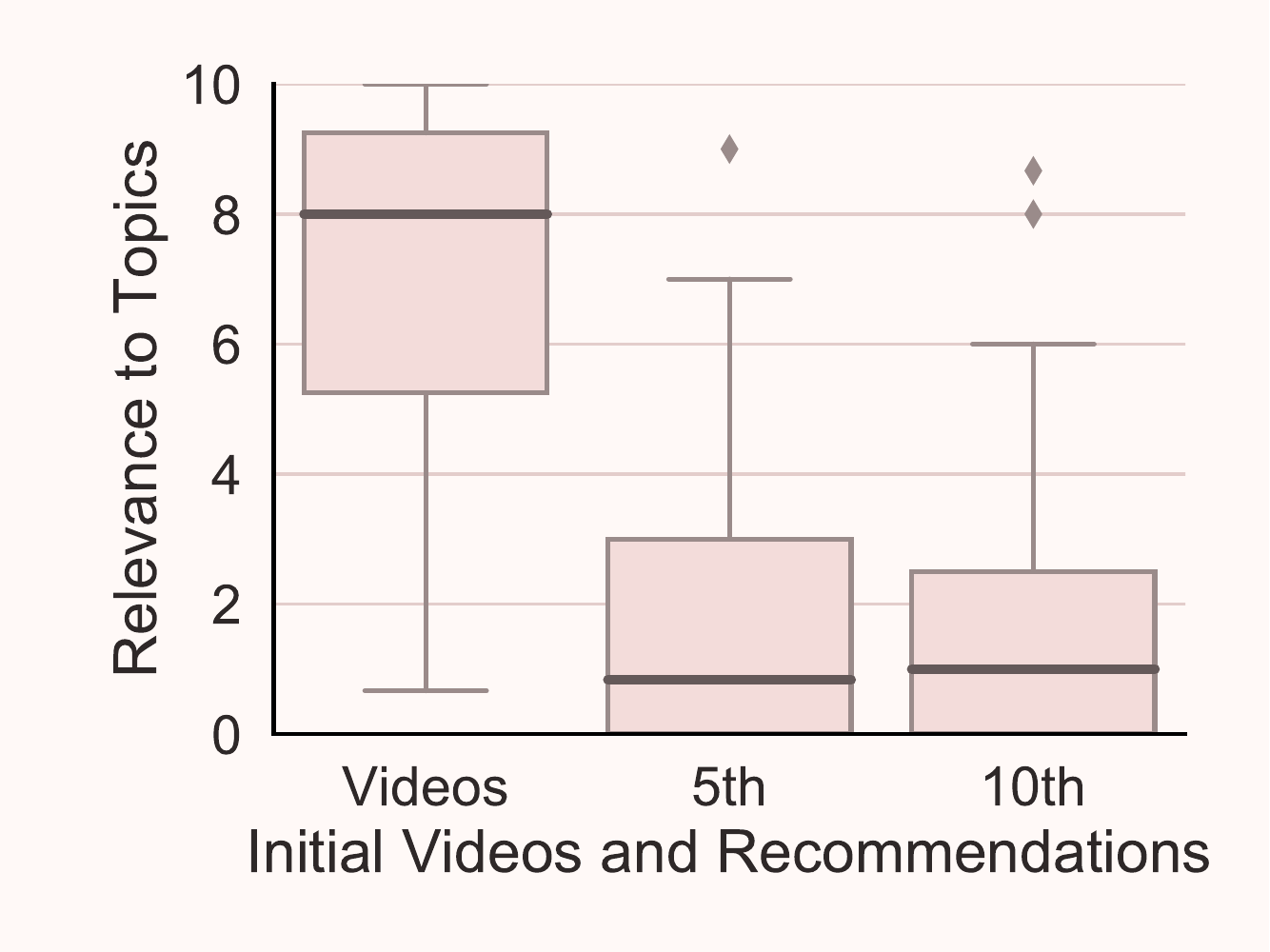}
 \caption{How related a video is to the nine topics in our investigation decreased significantly between the initial video and the 5th recommendations}\label{fig:relation_to_topic}
\end{figure}

The raters coded how relevant each video was to the list of topics that we provided. The results show that raters consider the initial videos to be very related to the political topics used as search terms. The median topic similarity rating of the initial videos was 8. This decreased dramatically to 0.83 after following only five recommendations. The similarity remains very low for the 10th recommendations, with a median rating of 1.00. Figure~\ref{fig:relation_to_topic} shows the boxplots of the ratings, whose interquartile ranges are decreasing. A two-tailed Mann-Whitney test indicated that the topics in the videos changed between the initial videos and the 5th recommendations ($U=607.5, p=.0000$), and between the initial videos and the 10th recommendations ($U=638.5, p=.0000$). These results indicate a strong topic drift. Recommended videos are about significantly different topics than the initial videos, which are based on the search results for the political topics that were entered into YouTube's search bar. This topic drift has important consequences for how suitable YouTube is as a provider of news.

\subsection{Emotions Evoked by the Recommendations}

\begin{figure}
\centering
 \includegraphics[width=.35\columnwidth]{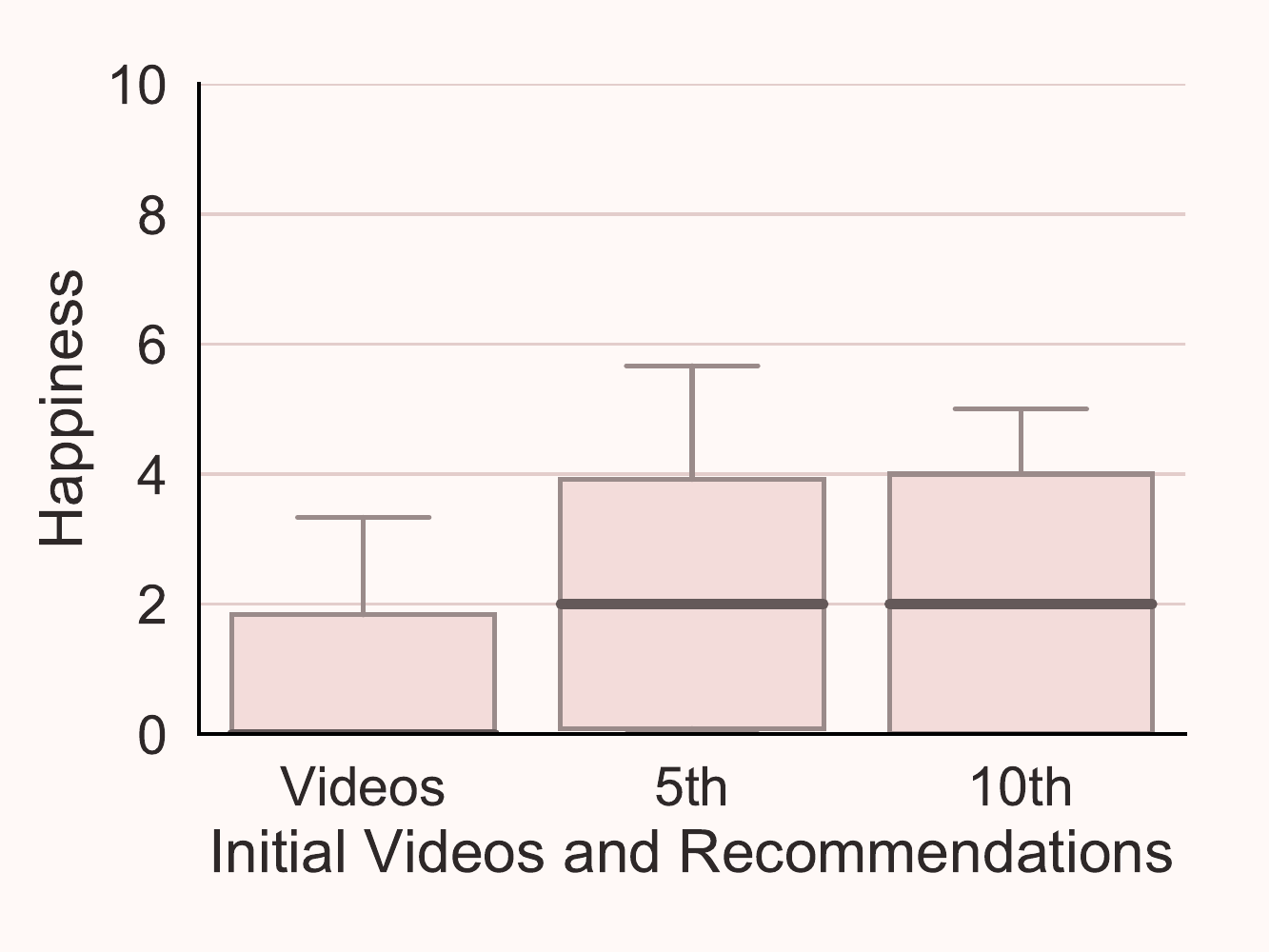}
 \caption{Boxplots of happiness evoked by the initial videos, the 5th recommendations and the 10th recommendations}\label{fig:emotion_boxplots_happy}
\end{figure}

\begin{figure}
\centering
 \includegraphics[width=.35\columnwidth]{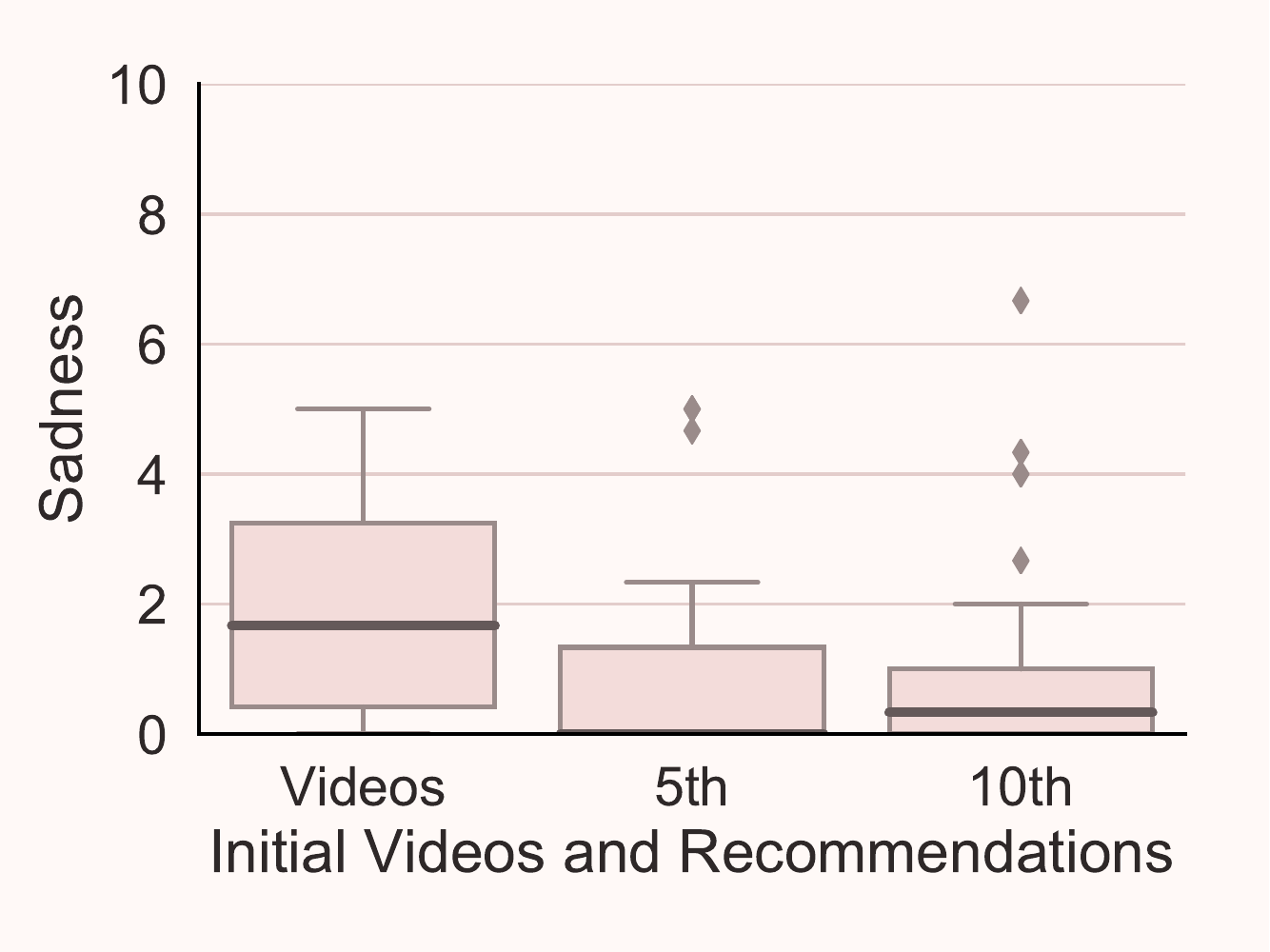}
 \caption{Boxplots of the sadness evoked by the initial videos, the 5th recommendations and the 10th recommendations}\label{fig:emotion_boxplots_sad}
\end{figure}

Finally, we investigated how the emotions evoked by the videos changed between recommendations. The raters evaluated whether the videos made them feel happy or sad on an 11-point Likert scale from \textit{``least''}~(0) to \textit{``most''}~(10). The goal of this was to investigate the relative change between the initial videos and the recommendations. Figure~\ref{fig:emotion_boxplots_happy} shows boxplots of the happiness evoked by the videos, which changes from a median of 0.00 for the initial videos to a median of 2.00 for the 5th and 10th recommendations. While 75\% of initial videos have a happiness rating between 0.00 and 2.00, more than half of the 5th recommendations and 10th recommendations have a happiness rating higher than 2.00. While the changes are small, two-tailed Mann-Whitney tests in Table~\ref{tab:emotion_mannwhitneyu} suggest that the differences between the initial videos and the recommendations are statistically significant. 

Regarding the sadness evoked by the videos, the trend is the opposite. The median ratings in the boxplots in Figure~\ref{fig:emotion_boxplots_sad} move from 1.67 for the initial videos down to 0.00~(5th) and 0.33~(10th). While more than half of the initial videos have a sadness rating higher than 1.67, 75\% of the 10th recommendations have a rating smaller than 1.00. The Mann Whitney~U tests in Table~\ref{tab:emotion_mannwhitneyu} show that these differences are unlikely due to chance. For happiness, all changes are significant at $p<0.01$. For sadness, the differences between the initial videos and the 5th recommendations are significant at $p<0.05$.

Our results show that the recommendations on YouTube are becoming increasingly more popular, more related to positive emotions, and less related to the initial political topics that we used as search terms.

\begin{table}
\centering
\caption{Two-tailed Mann–Whitney U tests show significant differences between the initial videos and the 5th recommendations and the initial videos and the 10th recommendations for sadness and happiness. At $p<.01$~(**) for happiness and $p<.05$~(*) for sadness.}~\label{tab:emotion_mannwhitneyu}
\begin{tabular}{l r l r l}
& \multicolumn{2}{c}{\textbf{5th Recomm.}} & \multicolumn{2}{c}{\textbf{10th Recomm.}} \\\cmidrule(r){2-3}\cmidrule(r){4-5}
\textit{Affect} & \textit{U} & \textit{p} & \textit{U} & \textit{p} \\
\midrule
Sadness & 470.5 & 0.0114 * & 472.5 & 0.0273 * \\
Happiness & 188.5 & 0.0043 ** & 193.5 & 0.0033 ** \\[.25\normalbaselineskip]
\end{tabular}
\end{table}

\section{Discussion}

In this paper, we describe an exemplary audit of YouTube's recommendations for political topics. This extends on prior work that investigated whether YouTube's recommendations are biased \cite{doi:10.1177/1354856517737222,doi:10.1080/19312458.2018.1479843}. As described, the research gap considering the bias in the recommendations on YouTube is especially problematic considering the specific requirements regarding fair and balanced reporting and the protection of minorities. Laws like the German Interstate Broadcasting Agreement require broadcasters to ``indicate a plurality of opinion'' and to report in a fair and balanced manner that takes minority views into account. \added{We investigate YouTube, a platform that has been accused of ``radicalizing'' users. In an Opinion piece in the New York Times, \citet{nytimes_youtube_radicalizer_2018} argued that: ``YouTube may be one of the most powerful radicalizing instruments of the 21st century.'' To thoroughly understand whether such radicalization on YouTube is taking place, research has to show: 1.~that YouTube is presenting users with increasingly extreme content, 2.~that this extreme content negatively affects their attitudes, 3.~that this affects their intentions, 4.~that this changes their behavior. This paper addresses 1. and focuses on the scenario observed in Chemnitz, where YouTube users who want to inform themselves about a new topic encounter extremist or hyperpartisan videos in a comparatively short watch session of ten videos or less. Considering the lack of established methods to measure how extreme some content is, we selected the popularity, relevance to topics, and the emotions happiness and sadness as proxies. Informed by a number of news articles \cite{isaac_facebook_2016,frenkel_facebook_2018,buzzfeed_las_vegas_2017,nytimes_chemnitz_2018}, we expected YouTube's recommendation system to zoom in on a topic (RQ2) by evoking strong negative emotions (RQ3) and by recommending less popular and more niche over time (RQ1) \cite{roose_youtubes_2019,silverman_viral_2016}. For the nine political issues from Germany, we showed that this is not the case. YouTube's recommendations enact a strong popularity bias (RQ1) and a noticeable emotionality bias (RQ3), without zooming in on a particular topic (RQ2)}. We found that YouTube's ML-based recommendation system has a strong tendency to recommend popular content. This is in line with prior work by Smith et al. \added{who performed more than 174,000 random walks and who analyzed more than 346,000 unique recommended videos \cite{pewresearch_youtube_2018}}. Considering some important methodological differences, our findings are especially noteworthy. \added{We corroborate Smith et al.'s findings in the context of German political topics and in a setting that is independent of YouTube's API. Our investigation was performed in German and used political topics as a starting point.} The random walks by Smith et al. exhibit a strong popularity bias since they used videos from the 14,000 most popular English-language YouTube channels (with at least 250,000 subscribers) as their starting point for the random walks. Unlike Smith et al., our investigation was based on the search results for current political topics. \added{We realistically model the situation in Chemnitz, where a user encounters a new topic for which he or she is trying to inform themselves. In addition to the views investigated by Smith et al., we also investigated the number of likes, which are a more explicit indicator of popularity \cite{lee2016makes}}. We found that the recommended videos that are based on search results are becoming increasingly more popular, both measured by how many times a video was viewed and by how many likes a video received. All differences considered, our results confirm that YouTube is guiding users towards increasingly more popular content. Our investigation indicates that one year after Smith et al.'s analysis, the popularity bias of the recommendation system had not changed. It is also noteworthy that the top search results, i.e. the initial videos that were returned for our searches, are significantly less popular than the recommendations. We think that this is surprising and invite further research to investigate this.

\subsection{Popularity Bias in YouTube's Recommendations}

YouTube markets its recommendation system as ``a sophisticated algorithm to match each viewer to the videos they are most likely to watch and enjoy'' \cite{youtube_creators_algorithm}. We found that popular, unrelated content is king. Our results indicate that while the first search results are relatively close to a given topic, the 5th recommendations are already very far from any of the topics in our investigation. \added{The increasing number of likes and views of the recommendations was surprising to us. We expected YouTube's search engine and its recommendation system to provide the most popular videos right from the start, i.e. in the search results and as the first recommendation. We did expect that the popularity of recommended videos would decrease over time. Instead, we found that the popularity of recommendations increased and that the topical relevance of the videos was low. In fact, our results show that the recommended videos are not even about political topics.}

We found no indication that people are consistently directed towards videos on a certain issue. This is surprising considering published work that indicates YouTube is or was based on association rule mining and that the videos a user watched are taken into account to recommend videos \cite{covington2016deep}. However, from a platform perspective, this makes sense. \added{For the platform, longer watch times could result in more ads that are shown, which could, in turn, lead to higher ad revenue}. It also connects to the investigation of the big data public by Harper \cite{doi:10.1177/1461444816642167} and the problem of the recentering of public engagement around the complementary interests of the broad majority and profitability. The audit showed that YouTube's recommendations are a prime example of this recentering. Moreover, the number of likes of the videos, which can be interpreted as a sign of virtual endorsement \cite{lee2016makes}, increased significantly. This means that the content presented by the recommendation system did not just show videos that people watched more, but content that a large group of people explicitly endorsed.

\subsection{Reporting Standards for ML-based Systems}

The popularity bias for political topics that we discovered is problematic from a democratic point of view. If the same laws that apply to private broadcasting services in Germany would be applied to YouTube's recommendation system, the popularity bias we discovered could violate Article 25 (1) of the German Interstate Broadcasting Agreement. The nationwide law for radio station and television licensing forces private broadcasters to present content that generally indicates a plurality of opinion. Our investigation showed that YouTube's recommendations for political topics are mainly focused on popular content that resonates with the majority.

The paper showed that the number of views and likes is increasing significantly for the recommendations. \added{While it can be assumed that the vast majority of relevant political, ideological, and social groups are represented in the vast amount of content on YouTube, their ML-based curation system is responsible for 70\% of the videos that users watch \citep{solsman_youtubes_2018}. The role of the ML-based curation system, therefore, requires special attention. We showed that the number of views and likes is increasing significantly for the recommendations. This could imply that popularity, as measured by likes and views, is the defining factor for selecting recommendations. If this is the case, then minority views are not adequately taken into account by the recommender systems. By definition, content aimed at a minority group is not able to get as many views or likes as content aimed at the majority group. If popularity, as measured in views and likes, is the defining factor for recommendations, minority groups are at a disadvantage. This is especially concerning for the political topics that we investigated.} Controversial political topics require a balanced presentation of all arguments in a way that weighs the pros and cons. The increase in popular, off-topic recommendations that we found suggests that YouTube's ML-based recommendation system is not suited to help users inform themselves about complex political issues. The popularity bias evidenced by the audit poses the question of whether all political, ideological, and social groups have adequate opportunity to express themselves in the \textit{``programme''}, i.e. the recommendations automatically provided by YouTube. This issue is not unique to YouTube but applies to all platforms where ML-based recommendation systems curate content. We would like to invite more researchers to investigate what can and should be expected from ML-based curation systems in this regard.

\subsection{From Popularity to Emotions}

Our study showed that the sadness decreased and the happiness increased in YouTube's recommendations, even for political topics. \added{This, again, was surprising to us, because we expected recommendations for political topics to evoke strong negative emotions to keep users engaged \cite{stieglitz2013emotions,ferrara2015quantifying}.} The most straightforward explanation for this would be that YouTube is actively optimizing its recommendations to increase happiness and to decrease sadness. While prior work showed that YouTube videos can be effectively classified into suitable emotion categories \cite{CHEN201740}, \added{based on what we have gathered about YouTube's recommendation system \cite{roose_youtubes_2019,covington2016deep}}, we think that it is highly unlikely that YouTube is actively optimizing recommendations for certain emotions. We do, however, think that the change in emotions in the recommendations could be explained by users regulating their emotions en masse using YouTube. Gross \cite{gross1998emerging} defined emotion regulation as: 

\begin{quote}
The process by which individuals influence which emotions they have, when they have them, and how they experience and express these emotions.
\end{quote}

This connects to early work by Bryant \& Zillmann \cite{bryant1984using}, who found that exciting or relaxing TV content is used to overcome boredom or stress. This also relates to the Mood Management Theory by Zillmann \cite{zillmann1988mood}, which states that the consumption of entertaining messages can alter mood states. We, therefore, believe that users could use YouTube to improve their mood by watching happy videos, which could influence the popularity signals that YouTube is relying on \cite{youtube_creators_home,covington2016deep}, thus increasing the happiness evoked by the platform. This is problematic considering the finding by Kramer et al. \cite{Kramer8788} that the emotions encountered on social media platforms like Facebook influence the emotions shared on Facebook.

\subsection{Auditing ML-Based Curation Systems}

Machine learning-based systems rely on data. An ML algorithm merely describes how the ML model is inferred from data. For this reason, ML-based systems cannot be studied using code audits that investigate the source code \cite{doi:10.1177/20539517211017593}. \added{This paper applies sock-puppet audits to scrutinize public relevance algorithms like YouTube's recommendation system}. As shown, the audits described in this paper allow stakeholders to examine the actions of ML-based systems. This paper applied audits to identify two important biases enacted by YouTube's recommendation system. One important benefit of such audits is that they can be conducted independently of the platform provider. \added{We describe how audits could enable researchers, non-governmental organizations, lawmakers, and other stakeholders to understand and monitor the recommendations of complex socio-technical systems like YouTube's recommendation system. Researchers could use audits to expose the biases enacted by the system. Therefore, the paper recommends audits as an important method to study algorithms}. While this paper focused on a particular snapshot of recommendations, the methodology can be adapted to audit systems over longer periods. The method could also be used to study phenomena like fake news and online radicalization. Audits are useful because they can be conducted ad-hoc and since they enable non-ML experts to identify biases enacted by an ML-based system. Prior work also suggests that audits are meaningful to laypeople \cite{pewresearch_youtube_2018}.

\subsection{Enforcing Laws}

\added{With this paper, we showed how audits can be used to detect systematic biases in recommendations. As described, laws like the German Interstate Broadcasting Agreement demand that the content of private broadcasters must generally indicate a plurality of opinion.} For the German political topics we investigated, our results showed that YouTube's recommendation system is enacting a strong popularity bias. \added{We believe that our audit-based approach could be adapted and generalized to examine whether there are other imbalances in the representation of other groups, e.g. based on gender, ethnicity, or sexual identity. Audits could then be used to ensure that recommendation systems are free of biases. New institutions, analogous to the German TÜV and Stiftung Warentest, could be founded to enforce laws like the Interstate Broadcasting Agreement, e.g. by monitoring the activity of the recommendations that billions of users are interacting with while consuming their news online. More broadly, such institutions could ensure that recommender systems and other AI systems act in the interest of the public good. To increase their trustworthiness, institutions that audit public relevance algorithms should be governed by public law, i.e., they should be independent and reliably financed. For institutions under public law that enforce laws like the Interstate Broadcasting Agreement, an expert consortium could define a set of biases that should not exist in the system, e.g. gender biases, ethnic biases, or popularity biases. Lawmakers could even use such audits to fine those who do not comply with laws like the German Interstate Broadcasting Agreement. Until these institutions are founded, we encourage researchers and civic hackers to fill the gap and to monitor the biases enacted by complex ML-based systems.}

\section{Limitations}

While the primary contribution of this paper is methodological, we also present findings on the biases enacted by YouTube's recommendations systems for a particular point in time and certain topics. The results are representative of how a particular political topic is presented to those who have never searched for the topic. However, since YouTube is known to personalize recommendations based on factors like a viewer's watching and searching history \cite{covington2016deep,youtube_creators_home}, we do not know what influence personalization has on recommendations. 

We performed Random Walks, a method that has been previously applied to study YouTube's recommendations. Smith et al.'s random walks were criticized as artificial because they relied on YouTube's API \cite{pewresearch_youtube_2018}. As described in the methods section, we mitigated this problem by relying on a Firefox-based bot.

\added{Rating the emotions in videos is a challenging task. To rate the emotionality in videos, we relied on scholars with a background in media and communication and social sciences, who reviewed and rated the videos. The three raters were briefed to only rate the emotions they perceived while watching the videos. They did not know about the goals of our investigation or our research questions. Due to the inherent subjectivity of experiencing emotion and the lack of established methods to reach intersubjective agreement on emotions, we used the three raters as a valid, but imperfect proxy. Our reliance on comparatively young research assistants limits the generalizability of the emotional assessment to other age groups. We do, however, only rely on these ratings for relative comparison, i.e. we do not make absolute statements about the emotionality of the content. Each rater introduced his or her personal bias regarding what their definition of happiness and sadness was}. Despite the complexity of affective content analysis in videos \cite{7024148,abrilian2005emotv1}, we found substantial to moderate agreement for the topic similarity, happiness, and sadness. Moreover, we do not make claims about the absolute emotional content but focus on the relative change in emotion that we showed with Mann-Whitney tests.

\section{Conclusion}

This paper investigated audits as a way to detect bias in ML-based curation systems like YouTube. The methodological contribution of this paper is showing how audits of ML-based systems can be used to investigate YouTube's recommendation system. The paper showed that audits are an important way of systematically examining recommendation systems that increasingly act as broadcasters. Researchers and lawmakers should use audits to ensure that public relevance algorithms like YouTube are acting in the interest of the public. Future work can use these insights to compare recommendations across different topics and countries. The paper identified a popularity bias in the recommendations and discusses its implications. \added{Our findings imply that YouTube's ML-based recommendation system is not suitable to help users learn about complex political issues.}

In the paper, we demonstrated that audits can yield insights into ML-based curation systems. \added{We hope that this paper can contribute to an ongoing socio-technological discussion about algorithmic bias and algorithmic experience.} The audits showed that YouTube's recommendations are increasingly focused on popular content that has more views and more likes than the initial videos. An analysis of emotions evoked by the videos showed that the content is becoming increasingly happy and less sad. Recommendations are also becoming less and less related to the political topics we investigated. We conclude that even for political topics, the recommendations by YouTube are mostly focused on keeping users watching. For the context that the paper investigated - a context particularly noteworthy considering the legal situation in Germany - the results clearly show that YouTube's recommendations are becoming systematically more popular and, at least to some degree, more focused on positive emotions and less focused on negative emotions. We invite other researchers and civic hackers to use audits as a method to ensure that complex ML-based systems like YouTube's recommendation system act in the interest of the public.

\begin{acks}
The work of Hendrik Heuer and Andreas Breiter was funded by the German Research Council (DFG) under project number 374666841, SFB 1342.
\end{acks}

\bibliographystyle{ACM-Reference-Format}
\bibliography{references}

\end{document}